\documentclass[twocolumn,twoside,slac_two]{revtex4}
\usepackage{graphicx}
\usepackage{fancyhdr}
\renewcommand{\headrulewidth}{0pt}
\renewcommand{\footrulewidth}{0pt}

\begin{document}

\title{The Nuclear Emulsion Technology and the Analysis of the OPERA Experiment Data}

%

\author{Tsutomu FUKUDA   \footnotesize on behalf of the OPERA collaboration}
\affiliation{F-lab, Nagoya University, Furo-cho, Chikusa-ku, Nagoya, 464-8602, JAPAN}

\begin{abstract}
 OPERA is an experiment that aims at detecting the appearance of $\nu_\tau$ in an almost 
pure $\nu_\mu$ beam (the CNGS neutrino beam) through oscillation. OPERA is a hybrid detector 
that associates nuclear emulsions to electronic detectors. The nuclear emulsion provides the 
resolution necessary to detect $\nu_\tau$ CC interactions. The first physics run started in July 
and ended in November 2008. In this presentation, the status of the emulsion technology and of 
the analysis of its data is reported.
\end{abstract}

\maketitle

\thispagestyle{fancy}


\section{Nuclear Emulsion}
Nuclear Emulsion is a special type of photographic emulsion made of AgBr microcrystals interspersed 
in a gel matrix.  A charged particle passing through such medium ionizes the crystals along its path 
and produces a latent image. Upon a chemical process, the development, the particle trajectory is 
materialized by a line of grains of metallic Ag (0.5$\sim$1 $\mu$m diameter); typically the grain 
density is about 30 grains / 100$\mu$m (Fig. 1). Nuclear emulsion is a sub-micron 3D tracking detector with 
a resolution of 0.3$\mu$m. In particle physics, the nuclear emulsion technology is notably reputed 
for the discovery of the pion~\cite{ref:pion} and of the charm particle in cosmic-ray~\cite{ref:charm} 
and for the first observation of the tau-neutrino~\cite{ref:donut}.

\begin{figure}[h]
\centering
\includegraphics[width=60mm]{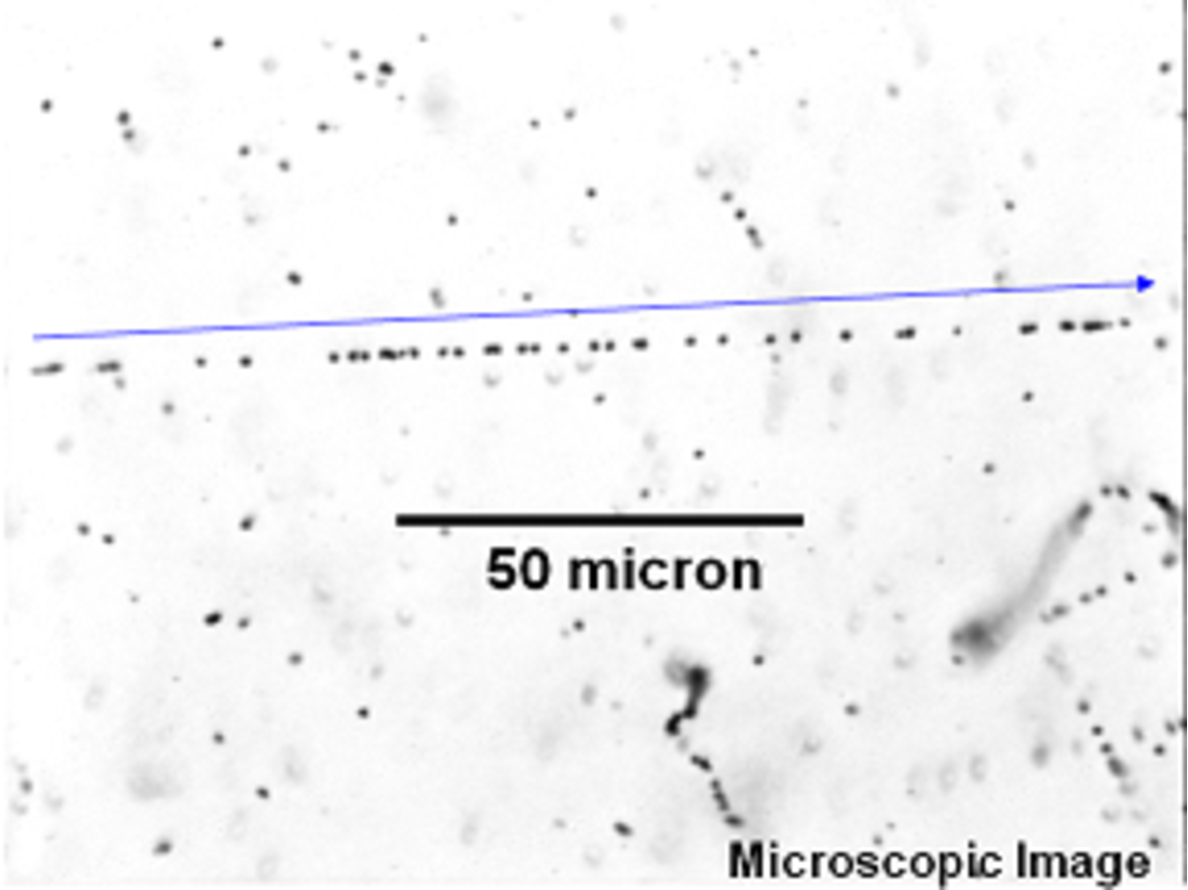}
\caption{A charged particle track in nuclear emulsion.} \label{figure01}
\end{figure}
\section{The OPERA Experiment}
In 1962, Maki, Nakagawa and Sakata proposed that oscillation may exist between massive neutrinos of 
different flavours~\cite{ref:mns}. In 1998, the Super-Kamiokande experiment established the deficit 
in atmospheric $\nu_\mu$ due to their disappearance through the oscillation mechanism~\cite{ref:sk}. 
This has been confirmed later by the K2K~\cite{ref:k2k} and then the MINOS~\cite{ref:minos} accelerator experiments. 
The goal of the OPERA experiment~\cite{ref:opera} is to detect for the first time the $\nu_\tau$ appearance 
in a $\nu_\mu$ beam in the atmospheric sector. The path length of the $\tau$ leptons produced in  $\nu_\tau$CC 
interactions is very short (c$\tau$= 87$\mu$m) thus requiring very high spatial resolution. In 2001, 
The DONuT experiment succeeded in detecting such interactions in Emulsion Cloud Chambers or ECC (Fig. 2). 

The detector is exposed to the CERN CNGS $\nu_\mu$ beam with an average energy of 17 GeV, well above the 
$\tau$ lepton production threshold in $\nu_\tau$ CC interactions~\cite{ref:cngs}. The rate of prompt $\nu_\tau$ 
is negligible at such energy. It is located in the Gran Sasso underground laboratory (LNGS) at a distance of 
730km from the neutrino source. For the most probable measured values of the oscillation parameters, $\Delta$ $m^2$ = 2.4 $\times$ 10$^-$$^3$eV$^2$, $sin^2$2$\theta$ = 1.0, 
the fraction of neutrinos having oscillated is about 1.7$\%$. The expected number of detected $\nu_\tau$ CC events 
is about 2.5 for a nominal year of run corresponding to 4.5~10$^1$$^9$ protons on target (pot). See also these proceedings~\cite{ref:serio}.

\begin{figure}[h]
\centering
\includegraphics[width=70mm]{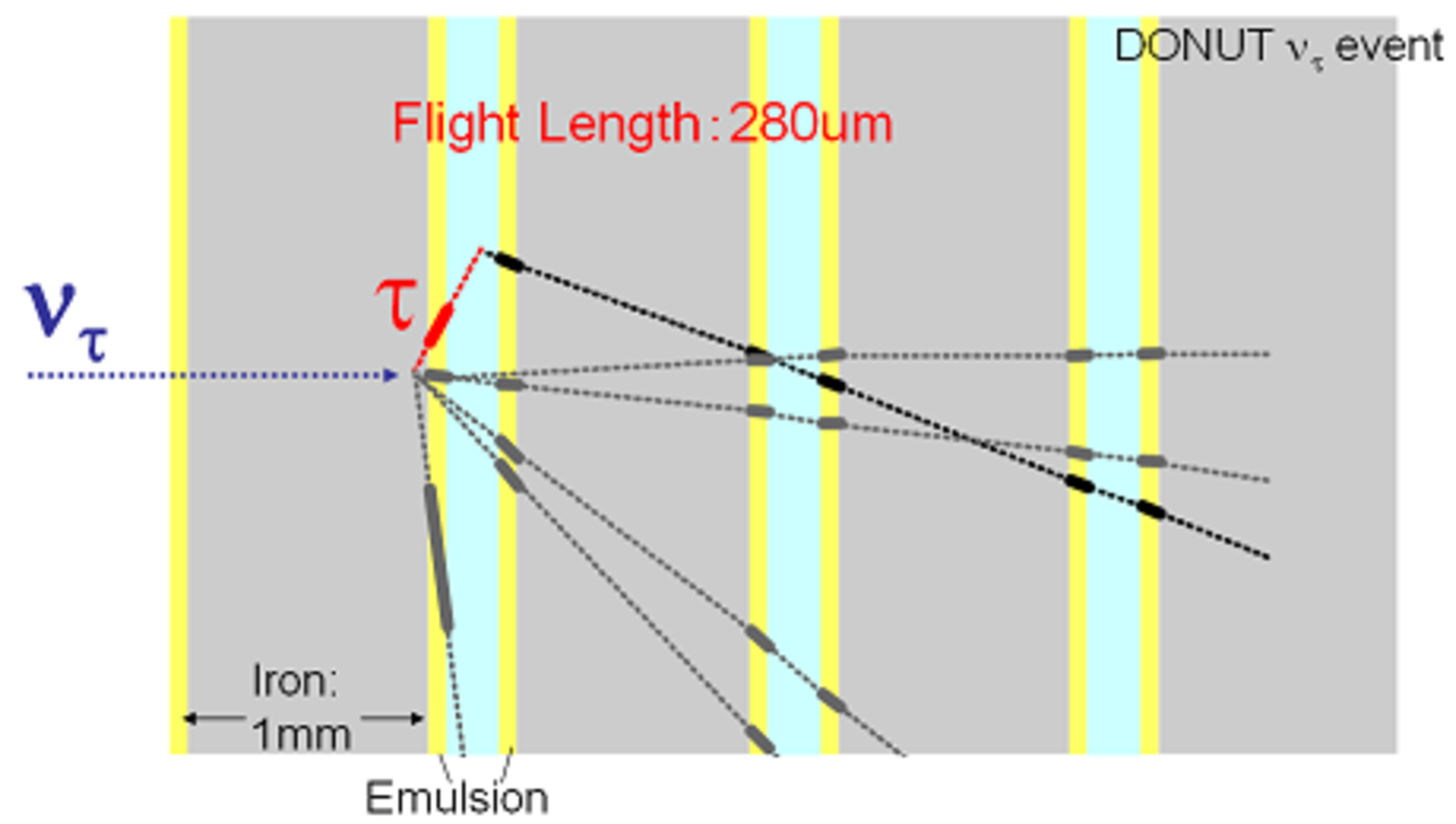}
\caption{A tau-neutrino event detected in the DONuT experiment.} \label{figure02}
\end{figure}

\section{The OPERA Detector}
OPERA is a hybrid detector composed of two identical super-modules. The targets, 1.25 kton in total, are each formed 
by about 75000 units called hereafter bricks based on the ECC technology. They are assembled into walls interleaved 
by two orthogonal planes of scintillator strips target trackers (TT)~\cite{ref:tt} used to identify the bricks in 
which the interactions occur.  Each target is complemented by a spectrometer that identifies muons and measures their 
charge and momentum. An overall picture of the detector is shown in Fig. 4 and its detailed description is available 
in ~\cite{ref:proposal}.

\begin{figure}[h]
\centering
\includegraphics[width=45mm]{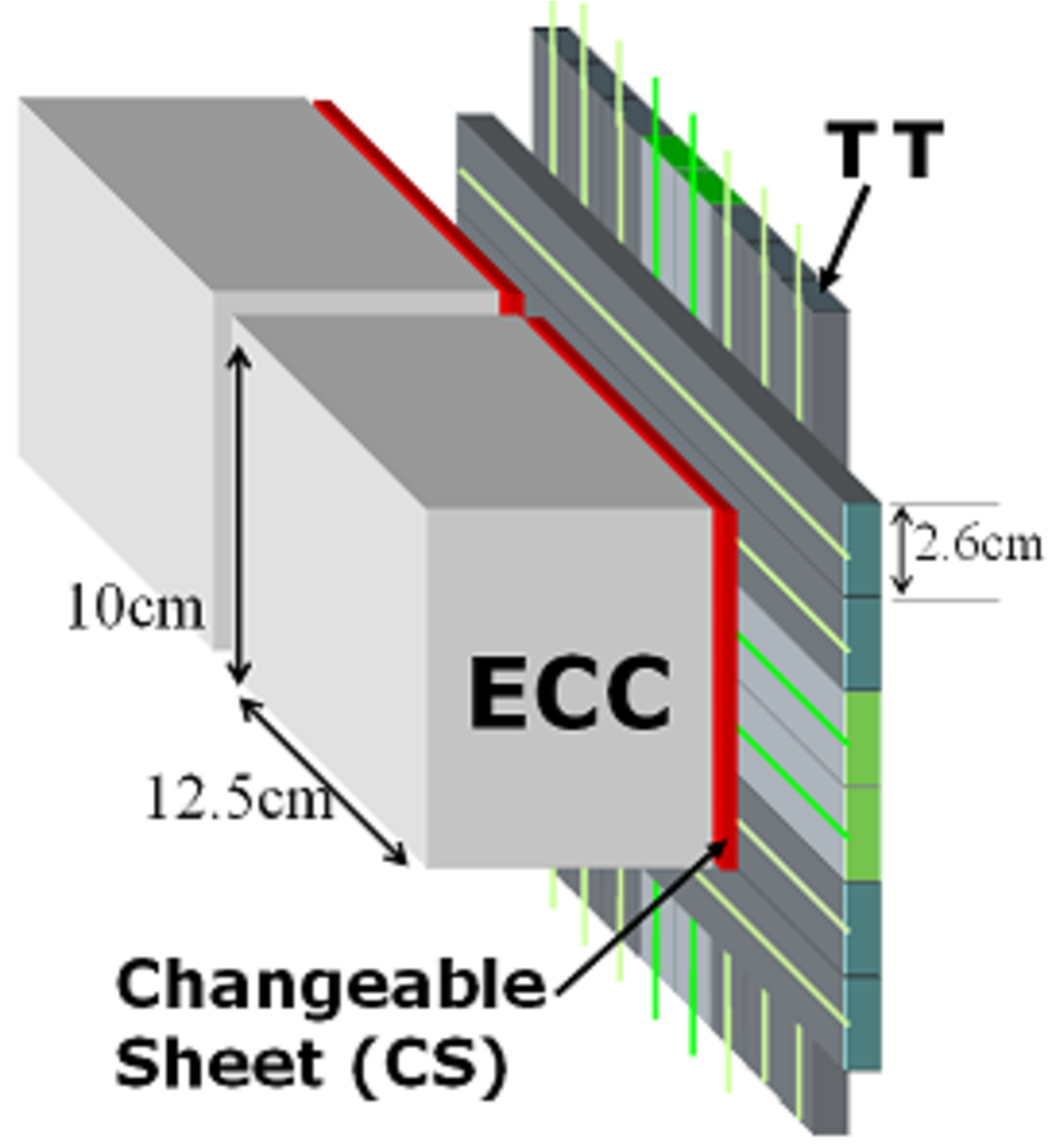}
\caption{ECC, CS, TT.} \label{figure03}
\end{figure}
\begin{figure}[h]
\centering
\includegraphics[width=75mm]{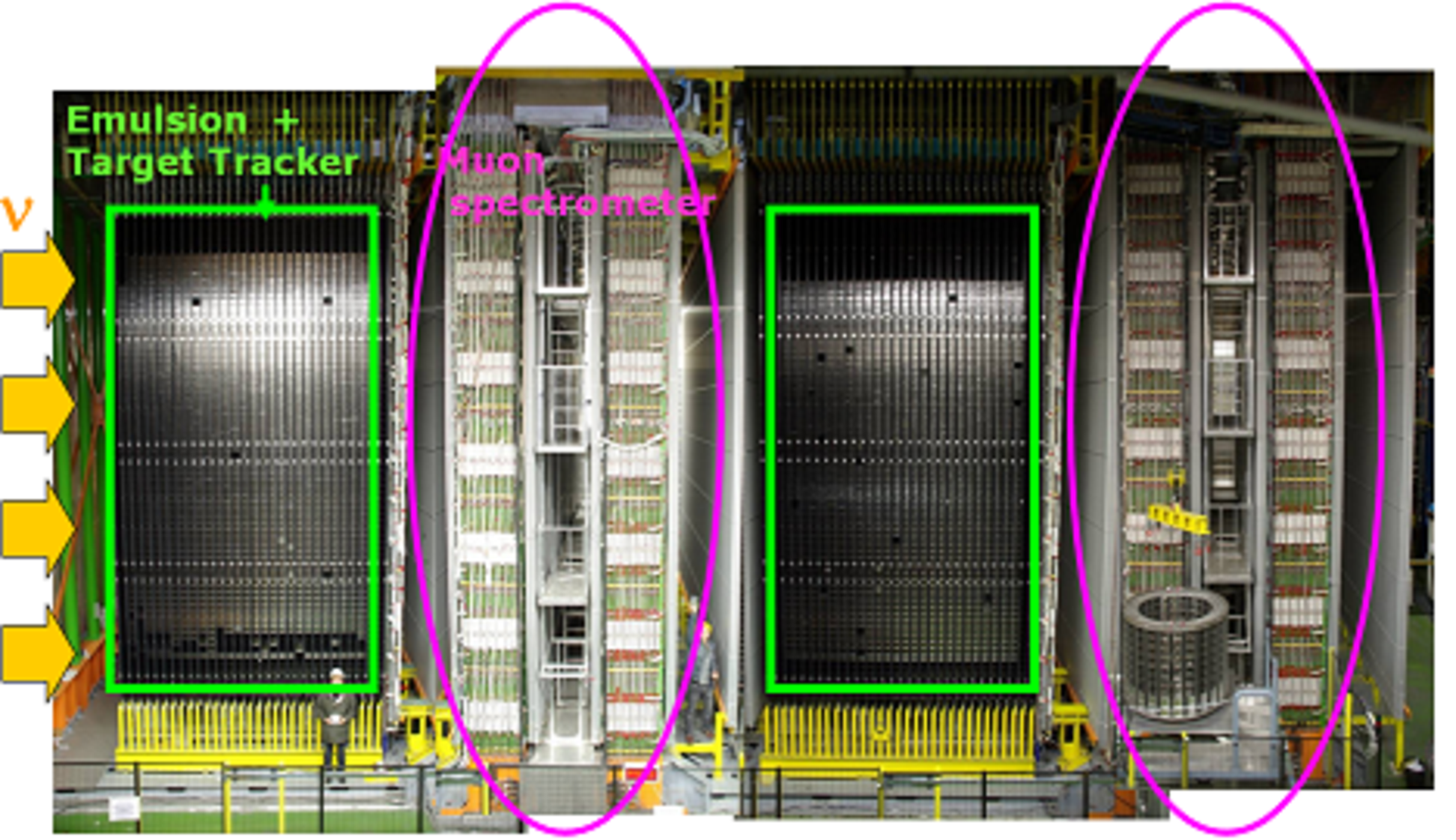}
\caption{The OPERA Detector @Gran Sasso (1400m underground).} \label{figure04}
\end{figure}

The size of a brick is 12.8cm $\times$ 10.2cm $\times$ 7.9cm. It is composed of 57 0.3mm-thick emulsion films~\cite{ref:emul} 
interleaved with 56 1mm-thick lead plates~\cite{ref:lead}. A film has a 44$\mu$m emulsion layer deposited on each side of a 205$\mu$m 
plastic base (Fig. 5). A separate box containing a pair of films hereafter called changeable sheets or CS is glued on the 
downstream face of each ECC brick in front of the next TT plane (Fig. 3). They serve as interface between the brick and the TT, 
bringing the centimetre spatial resolution of the TT down to the $\mu$m level.

\begin{figure}[h]
\centering
\includegraphics[width=80mm]{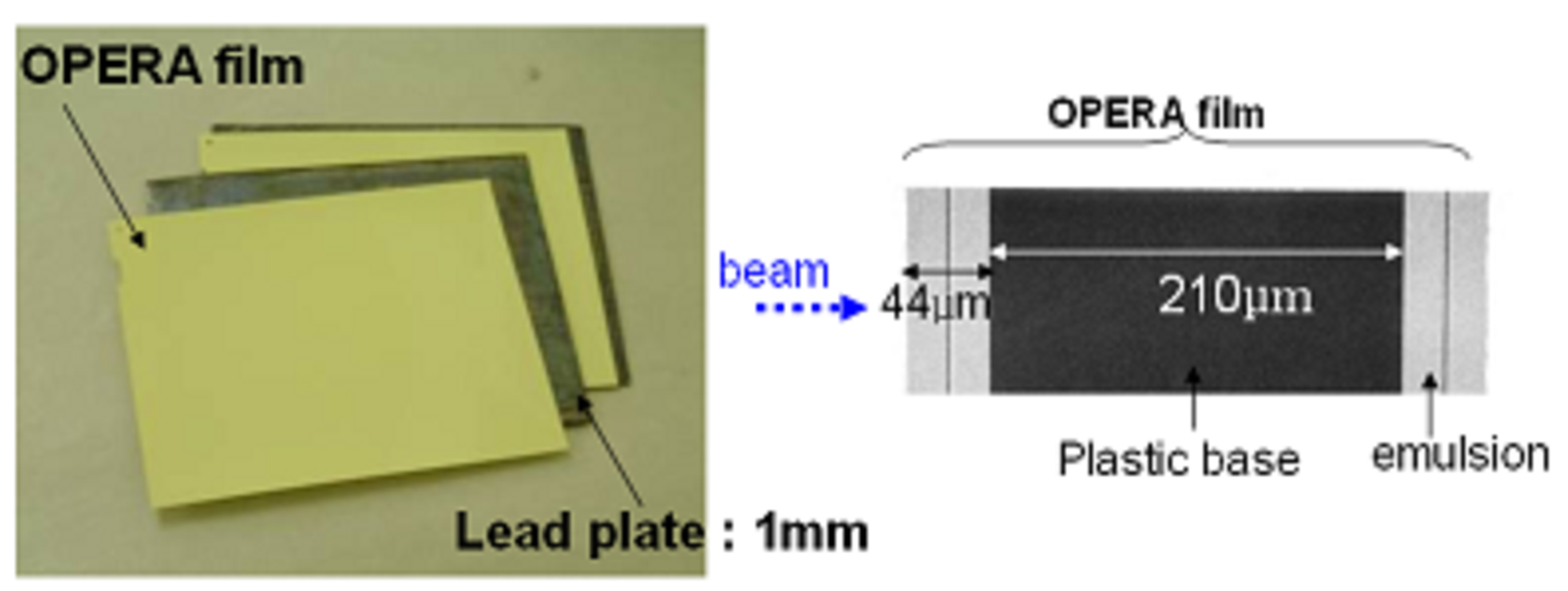}
\caption{OPERA film.} \label{figure05}
\end{figure}

\section{Flow of the event analysis}
The event analysis is performed in two main steps: location of the neutrino interaction and search for a secondary vertex 
topology of which kinematics is compatible with that of $\tau$ decay.

\subsection{Neutrino interactions location}
The signal recorded by the electronic detectors, TT and spectrometers, is used to identify the bricks, most often one and 
up to 3, in which the neutrino interaction is likely to have occurred. Those bricks are extracted from the target by an automaton. 
The CS are removed from their box, developed and analysed, starting with the most probable brick. The level of background 
in the CS is negligible; essentially only tracks from the neutrino interaction are recorded. The identification of the brick 
is thus confirmed by finding tracks in the CS that are compatible with the electronic data, in which case the brick is 
disassembled and its films developed. The tracks found in the CS are extrapolated to the most downstream emulsion film where 
they are searched for. They are then followed back from film to film up to the location of the neutrino vertex where they 
disappear. Finally emulsion data is taken around this point and the neutrino interaction vertex is reconstructed.

\subsection{Decay Search}
The four main decay channels of the $\tau$ lepton are given in Table I. Topologically, they are classified as "kink" or 
"trident" if they have one or 3 charged daughters. Charm particles have similar lifetime and decay topologies as the $\tau$ lepton. 
Understanding their detection rate is therefore a direct verification of the expected detection efficiency of the $\tau$ lepton.

\begin{table}[h]
\begin{center}
\caption{$\tau$ decay modes.}
\begin{tabular}{|l|c|c|c|}
\hline \textbf{topology} & \textbf{decay mode} & \textbf{ratio}
\\
\hline kink & $\tau$$\to$e & $\sim$18$\%$ \\
\hline kink & $\tau$$\to$$\mu$ & $\sim$17$\%$ \\
\hline kink & $\tau$$\to$hadron & $\sim$49$\%$ \\
\hline trident & $\tau$$\to$3hadrons & $\sim$15$\%$ \\
\hline
\end{tabular}
\label{table01}
\end{center}
\end{table}

Single prong events fall into three categories. For 60$\%$ of the events (Fig. 6-bottom), the decay occurs inside the vertex lead plate and 
the parent traverses no emulsion layer. The decay products are identified by their large impact parameter (IP) with respect 
to the primary vertex.  For 30$\%$ of the events (Fig. 6-top), the parent traverses at least one film. In this case, the trajectory of 
both the parent and the decay product may be reconstructed inside at least one plastic base from data registered in the two 
emulsion layers, the two micro-tracks, to form a base-track. The candidates are identified by the observation of a kink 
between both trajectories. For 10$\%$ of the events, the parent traverses only one emulsion layer and decays in the film base. 
There is no base track on the parent trajectory but a single micro-track. Whether the presence of this micro-track will is 
sufficient to identify the candidates by the observation of a kink or will it be identified by the IP method is under study.

\begin{figure}[h]
\centering
\includegraphics[width=60mm]{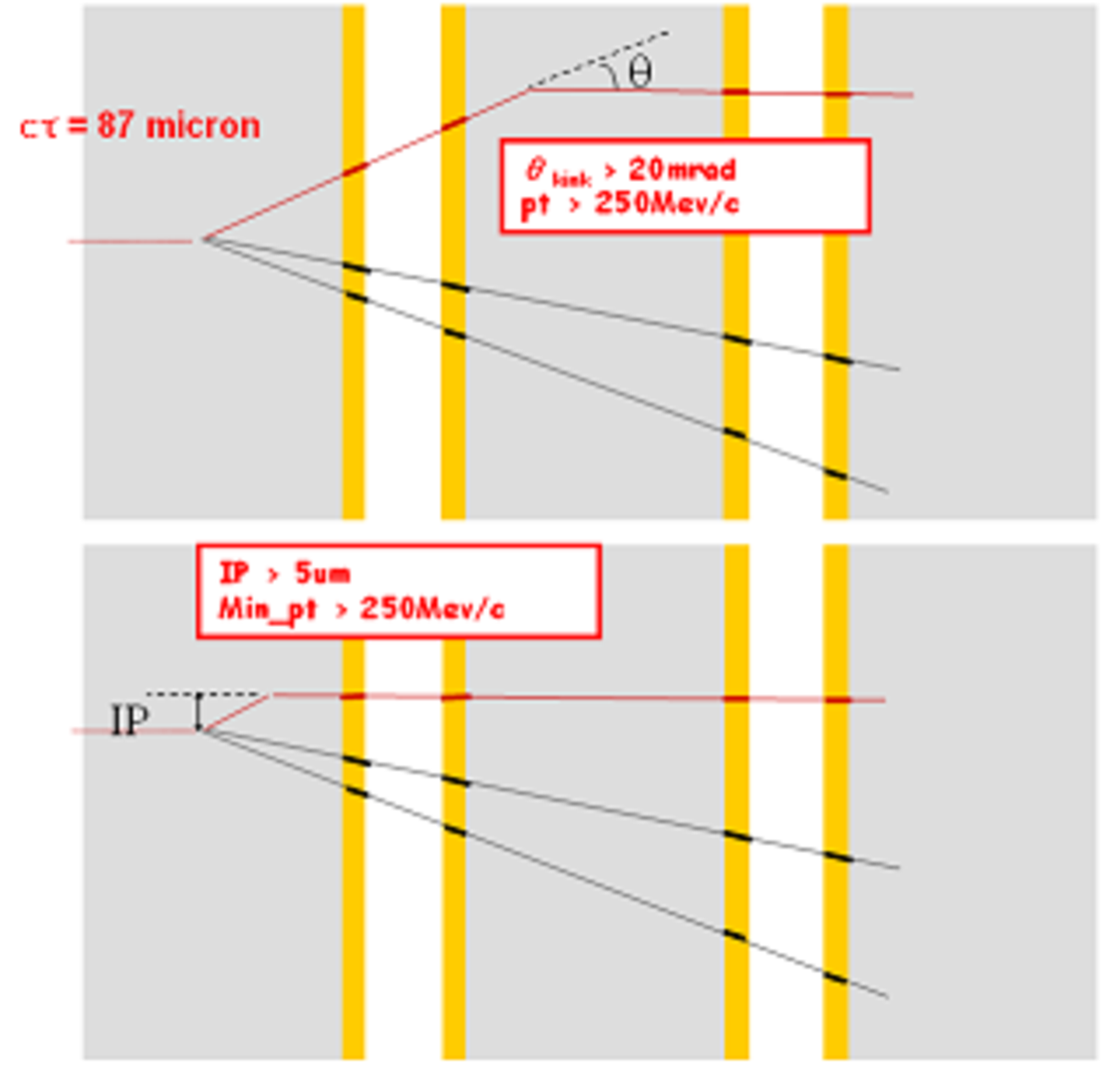}
\caption{Decay Search.} \label{figure06}
\end{figure}

\section{Emulsion Technology developed for the OPERA experiment}
\subsection{The OPERA film}
The mass production of the nuclear emulsion films used for OPERA is the first example of an industrial process. In the 
past history, nuclear emulsion film or plates were poured by hand. Machine production guarantees homogeneity in thickness 
and sensitivity unreachable before. An added protective coat allows hand manipulation of the film. A new feature of the 
OPERA film is that tracks already recorded may be erased. This process is called "refreshing".
\subsection{Film refreshing}
The "refreshing" process was developed to allow erasing the large background of cosmic ray tracks recorded in the films 
since their time of fabrication. The signal recorded in nuclear emulsion was known to fade with time at a speed depending 
on the environment, causing the sensitivity to track detection to progressively decrease. In the OPERA film, this fading 
effect is controllable. By keeping films at high relative humidity (98$\%$) and high temperature (30$^{\circ}$C) for 3 days, 
more than 99$\%$ of the recorded tracks were erased (Fig. 7) while the sensitivity to new recording is not affected. All 
the films were refreshed underground in the Tono mine in Japan and then transported to Gran Sasso at see level. The films 
used for the CS require very low background; they were refreshed a second time underground at Gran Sasso.
\begin{figure}[h]
\centering
\includegraphics[width=75mm]{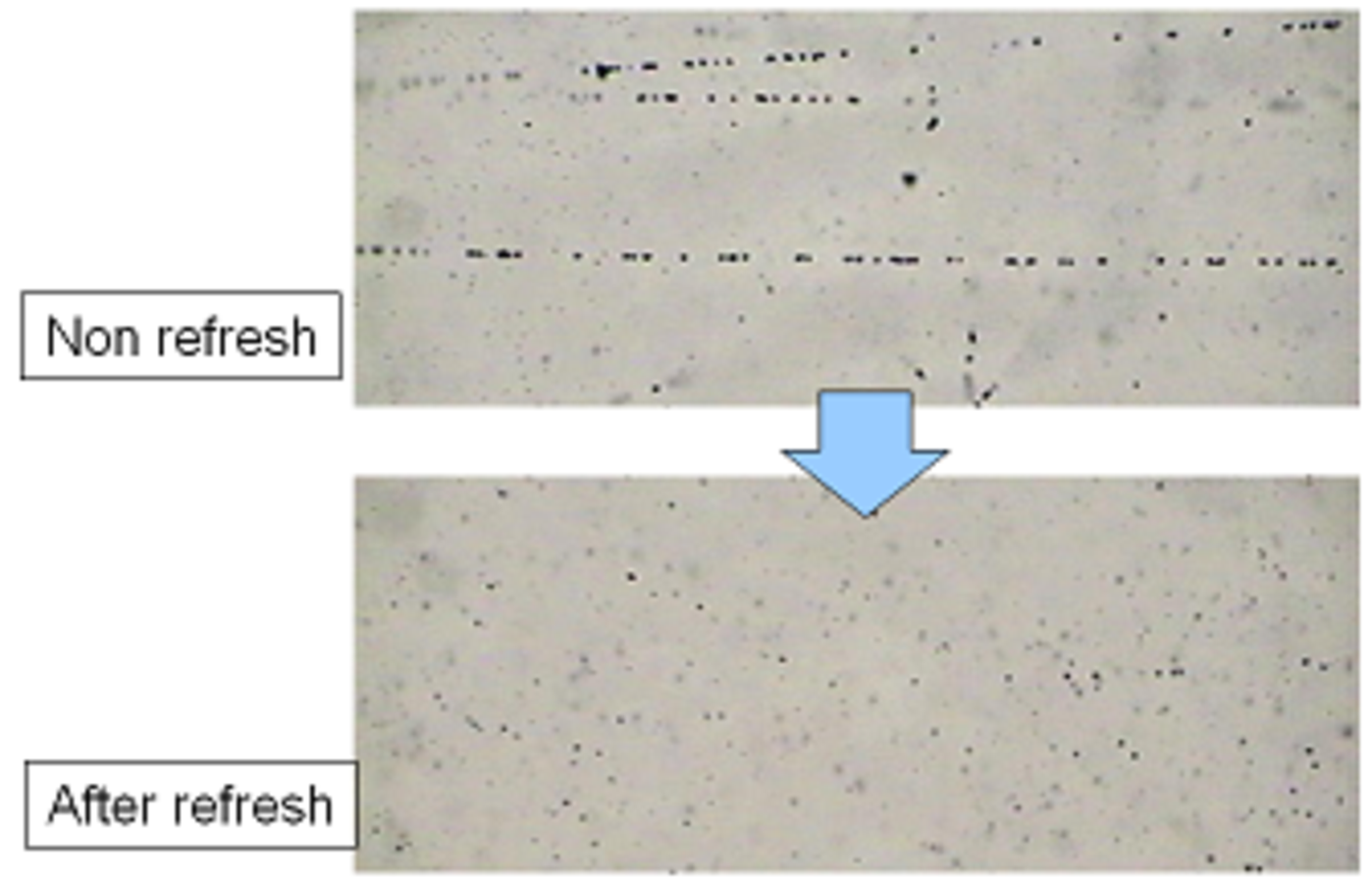}
\caption{Film refreshing.} \label{figure07}
\end{figure}
\subsection{High speed scanning systems}
The scanning area of the OPERA CS measures in cm$^2$; this is more than 100 times larger than for DONuT, the last experiment 
having used the ECC technique. Therefore high speed scanning systems were developed in both Japan and Europe. The new Japanese 
scanning system, the S-UTS (Super Ultra Track Selector) is shown on Fig. 8-left. Four systems with a scanning speed of 75cm$^2$ 
per hour and one of 20cm$^2$ per hour are operational in Japan. The previous generation operated at speeds of about 1cm$^2$ 
per hour~\cite{ref:uts}; 5 such systems are used for manual verification and subsidiary tasks. The European scanning machine~\cite{ref:ess}, 
the ESS (European Scanning System) is shown on Fig. 8-right. It operates at a speed of 20cm$^2$ per hour. A total of 33 such systems 
are currently active for OPERA.
\begin{figure}[h]
\centering
\includegraphics[width=80mm]{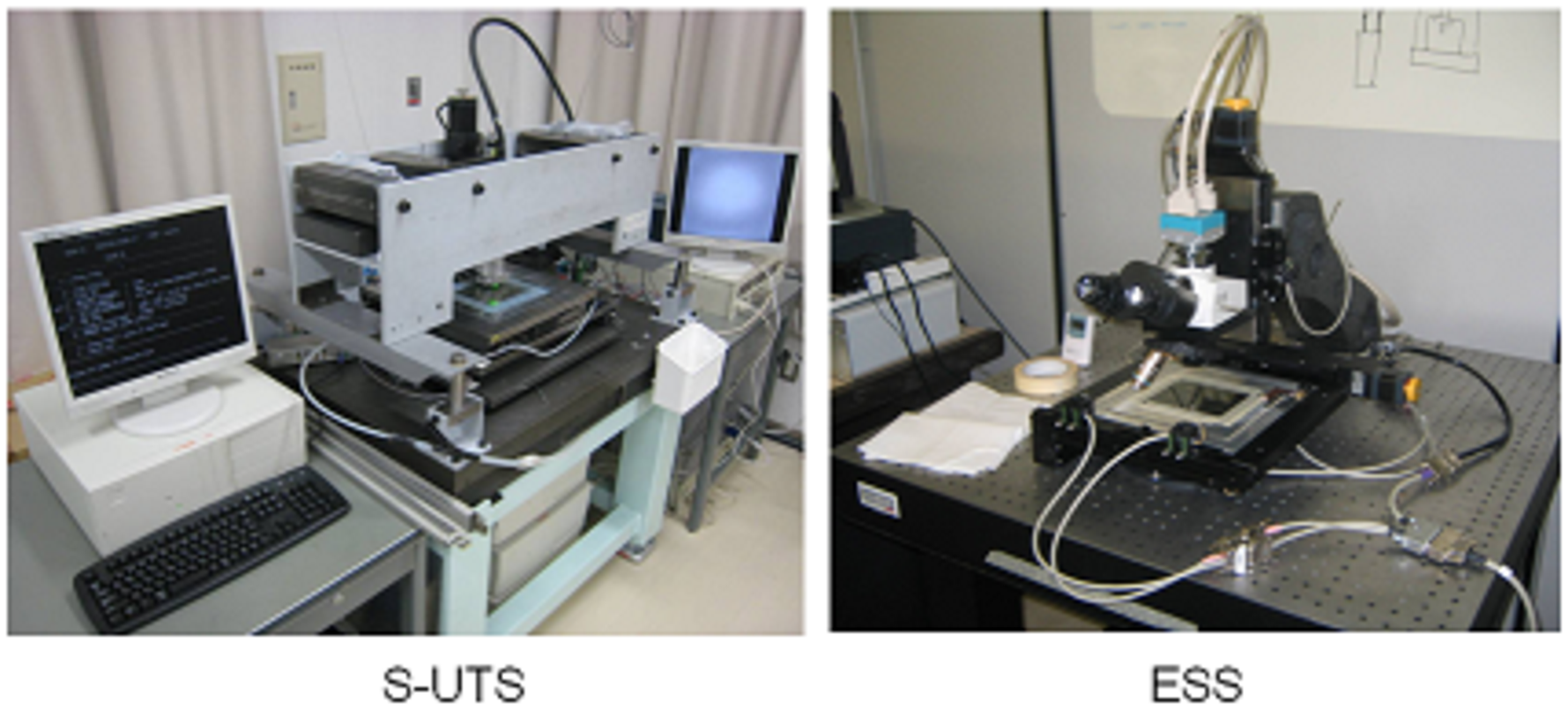}
\caption{New Scanning Machines.} \label{figure08}
\end{figure}
\subsection{Momentum measurement in ECC bricks}
The momentum of charged particles can be estimated in ECC bricks by measuring their Multiple Coulomb Scattering~\cite{ref:park}. 
When a particle of charge $z$, momentum $p$ and velocity $\beta$$c$ traverses a material of depth $x$ and radiation length $X_0$, 
the distribution of the scattering angle is expressed by a Gaussian, the RMS of which is approximately given by Eq.1.

\begin{equation}
\theta_0 = \frac{13.6MeV}{\beta cp}z\sqrt{\frac{x}{X_0}}
\label{eq-mcs}
\end{equation}

The measurement of the angle differences in two consecutive films provides the momentum estimation (Fig. 9). The results of a 
test experiment at KEK are shown in Fig. 10. A brick was exposed to 0.8GeV/c and 1.5GeV/c pion beams. The values of the momentum 
measured by MCS are respectively of 0.79GeV/c ($\sigma$p/p =11$\%$) and 1.53GeV/c ($\sigma$p/p = 16$\%$).

\begin{figure}[h]
\centering
\includegraphics[width=70mm]{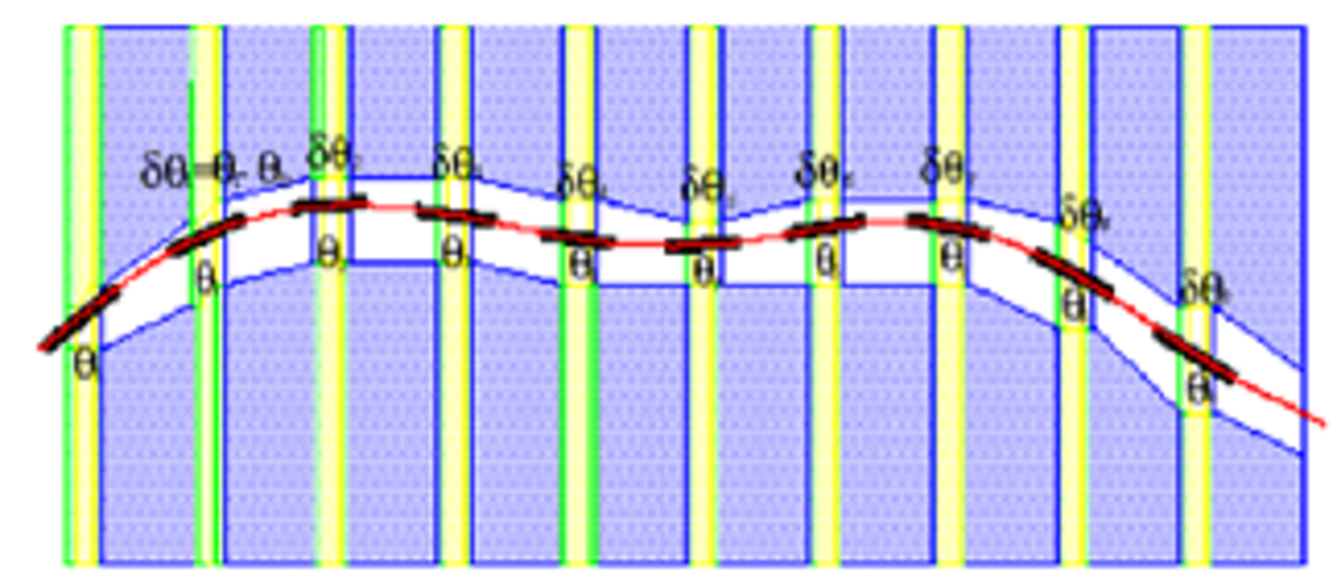}
\caption{Momentum measurement in ECC bricks.} \label{figure09}
\end{figure}
\begin{figure}[h]
\centering
\includegraphics[width=75mm]{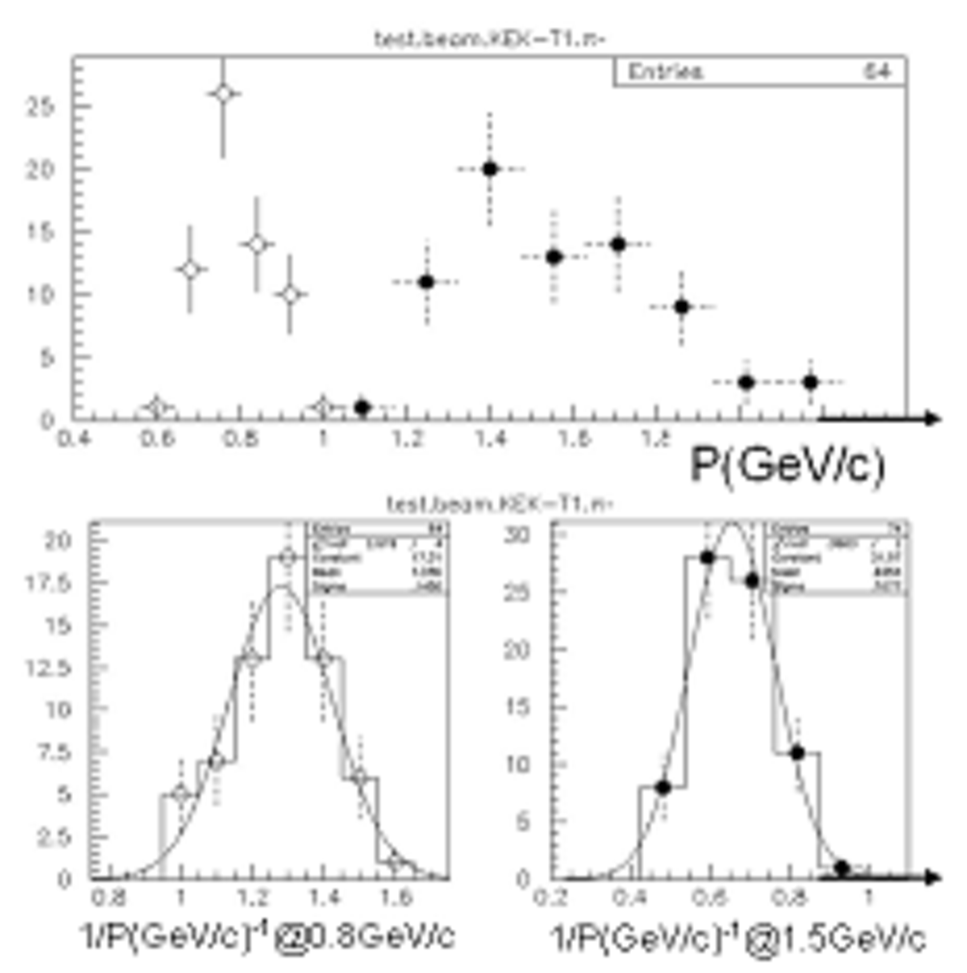}
\caption{Result of a test experiment at KEK.} \label{figure10}
\end{figure}

\subsection{Electron energy measurement in ECC bricks}
Fig. 11 shows the development of electromagnetic showers generated by electrons in an ECC brick. Counting track segments 
in the shower provides an estimation of the incident electron energy~\cite{ref:shower}. Fig. 12 shows the result of a 
test experiment at CERN with 2GeV and 4GeV electron beams. There is agreement between the measured and Monte Carlo 
simulated distributions of the numbers of segments though errors on data are large.

\begin{figure}[h]
\centering
\includegraphics[width=90mm]{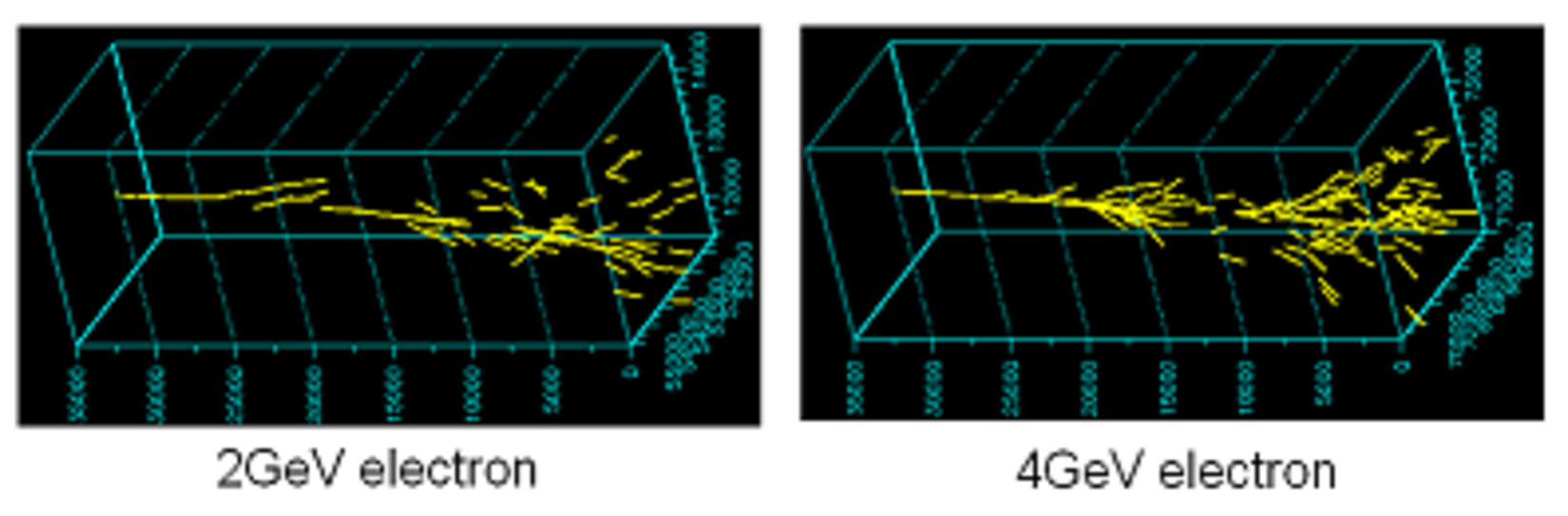}
\caption{Electron energy measurement in ECC bricks.} \label{figure11}
\end{figure}
\begin{figure}[h]
\centering
\includegraphics[width=75mm]{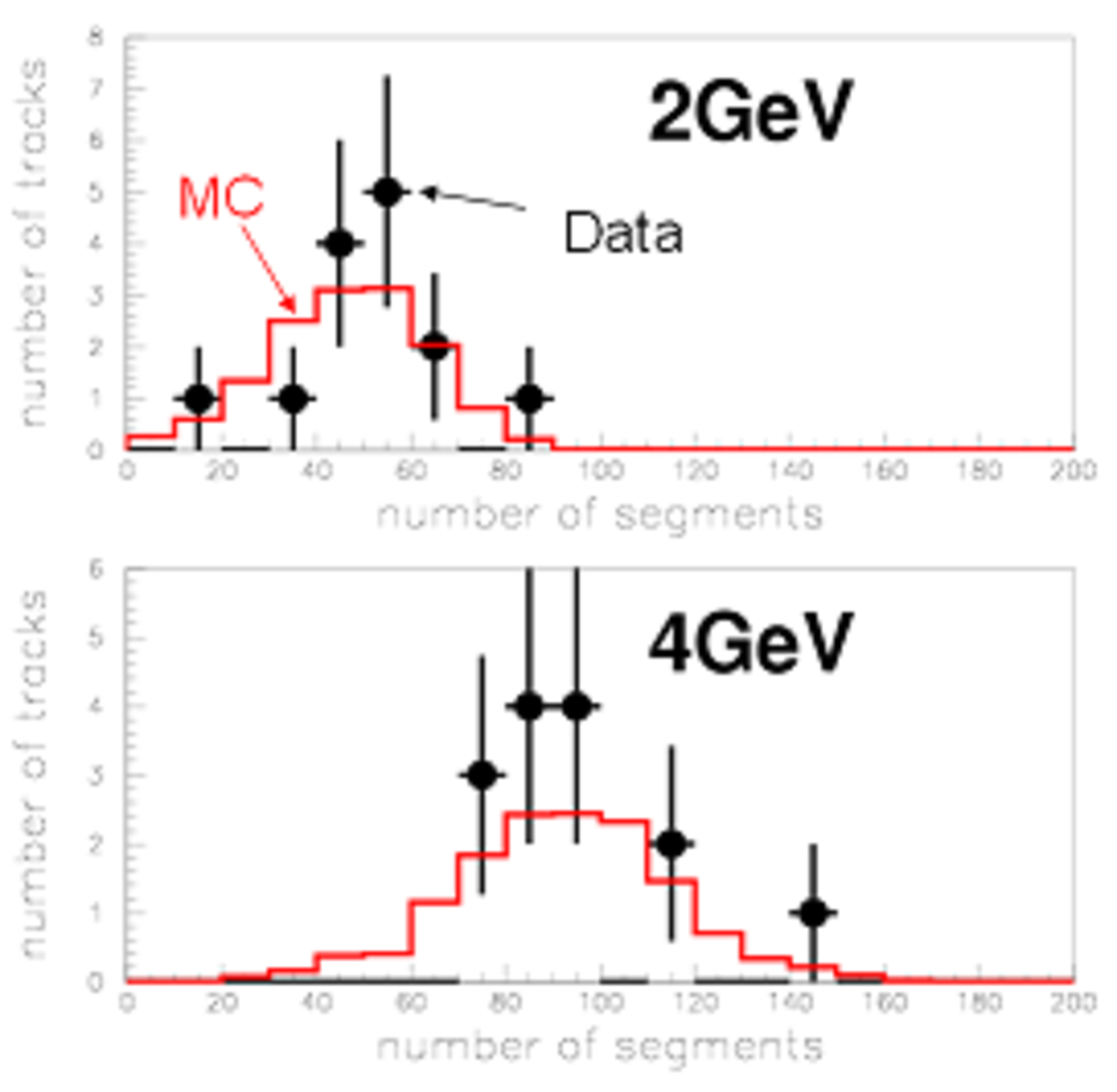}
\caption{Result of a test experiment at CERN.} \label{figure12}
\end{figure}

\subsection{dE/dX measurement in ECC bricks}
In nuclear emulsion, the grain density recorded along the track of a charged particle is almost proportional to its 
energy loss dE/dX (Fig. 13,~\cite{ref:ppi}). ECC bricks have been exposed to 0.4GeV/c, 0.5GeV/c, 0.6GeV/c, 0.74GeV/c, 
0.87GeV/c, 1.14GeV/c and 2.0GeV/c proton and pion beams at KEK. Fig. 14-left illustrates the particle identification 
capability of the method (pion, proton and deuteron contamination) at 0.87GeV/c. The relation between the grain density 
and the momentum of the particles is shown Fig. 14-right.

\begin{figure}[h]
\centering
\vspace{0.5cm}
\includegraphics[width=75mm]{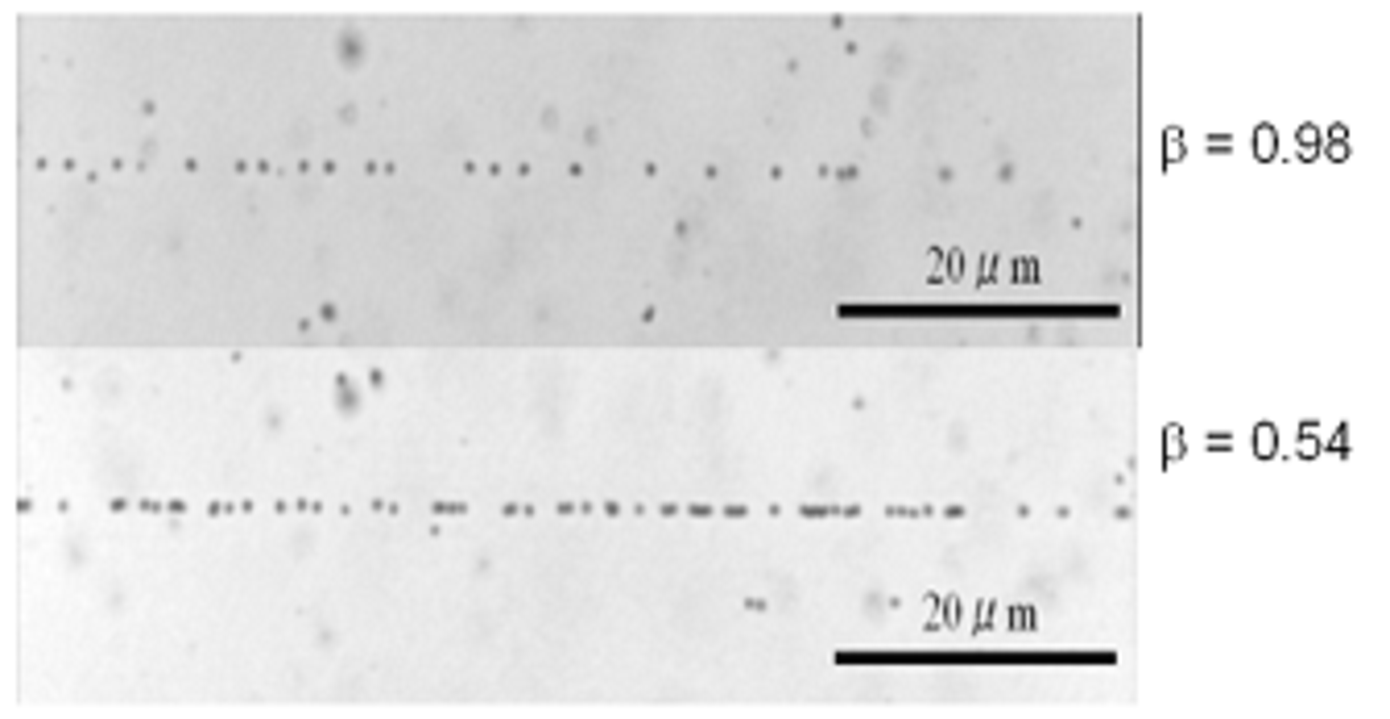}
\caption{A pion track of 0.6GeV/c(top) and a proton track of 0.6GeV/c(bottom).} \label{figure13}
\end{figure}
\begin{figure}[h]
\centering
\includegraphics[width=85mm]{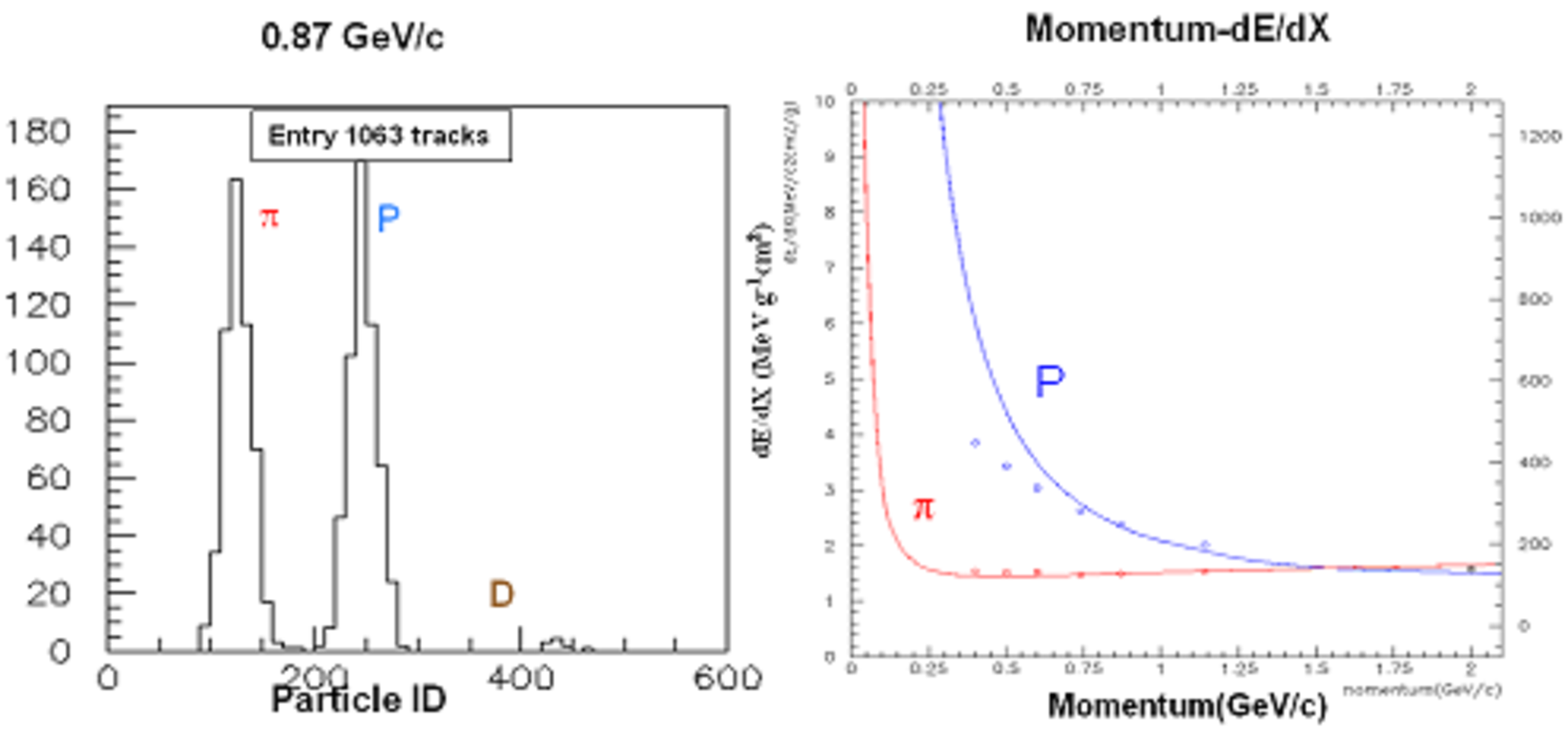}
\caption{Result of a test experiment at KEK.} \label{figure14}
\end{figure}

\section{Status of the 2008 OPERA run}

\subsection{Location status}
Triggers for a total of 1690 events in the target were generated by the TT in time with the CNGS beam arrival. Tables II and III 
summarize the current status of the events location in the CS and in the bricks after the analysis of the most probable 
CS indicated by the TT is well advanced. 797 events have been located already in the bricks. For events not found in the CS, 
the analysis of the next most probable CS has started.
\begin{table}[h]
\begin{center}
\caption{Current status of the event location in CS.}
\begin{tabular}{|l|c|c|c|}
\hline number of events & 1473 \\
\hline Scanning done & 1440 \\
\hline Found in CS & 1110 \\
\hline Next CS : now in progress & 330 \\
\hline
\end{tabular}
\label{table02}
\end{center}
\end{table}
\begin{table}[h]
\begin{center}
\caption{Current status of the event location in ECC bricks.}
\begin{tabular}{|l|c|c|c|}
\hline  & NC & CC & TOTAL \\
\hline ECCs received in laboratories & 218 & 959 & 1177 \\
\hline ECCs measured & 195 & 895 & 1090 \\
\hline CS to ECC connection successful & 178 & 849 & 1027 \\
\hline Neutrino interactions located & 119 & 678 & 797 \\
in the ECC & & & \\
\hline Neutrino interactions in the & 12 & 46 & 58 \\
upstream wall & & & \\
\hline Neutrino interactions in the & 4 & 17 & 21 \\
dead material & & & \\
\hline
\end{tabular}
\label{table03}
\end{center}
\end{table}
\subsection{Decay search status}
The search for decay vertices in the events already located is in progress. Fig. 15 shows in red the minimum distance 
between pairs of tracks with P$>$1GeV/c emitted in real neutrino interactions. In blue, it shows the Monte Carlo distribution 
of the IP with respect to the primary vertex of the daughter particle of $\tau$ leptons decaying in the lead plate of the 
primary vertex. The minimum distance between all pairs of primary tracks is within 10$\mu$m. This demonstrates that the 
OPERA ECC bricks have enough resolution to identify $\tau$ decay topologies based on the IP. 

\begin{figure}[h]
\centering
\vspace{0.5cm}
\includegraphics[width=80mm]{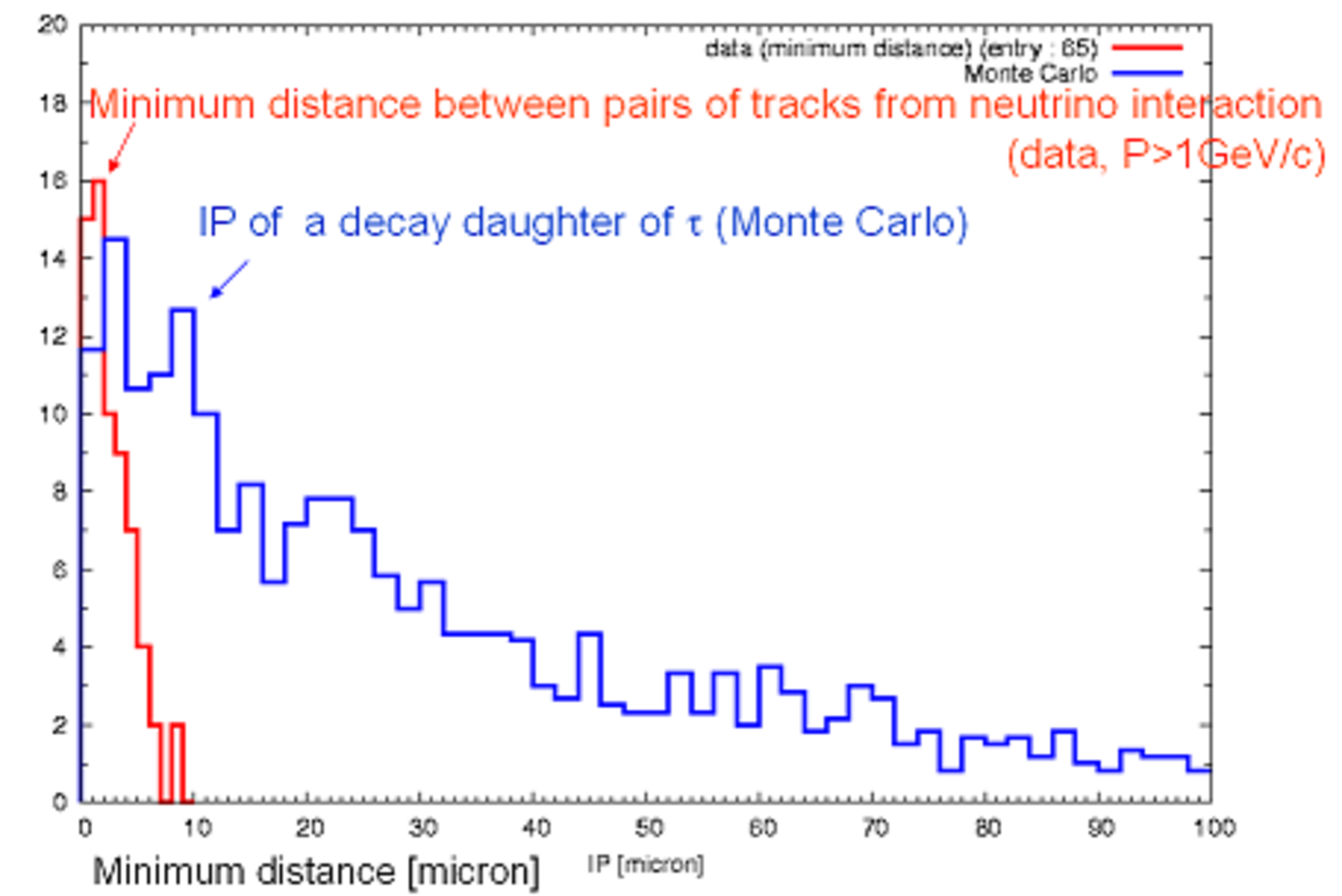}
\caption{Distribution of minimum distances between pairs of tracks with p $>$ 1GeV/c emitted in real neutrino interactions(red) and Monte Carlo 
distribution of the IP w.r.t. the primary vertex of the daughter particle of tau leptons(blue).} \label{figure15}
\end{figure}

Some charm decay candidates have already been detected as exemplified in Fig. 16. There are 6 tracks including a $\mu^-$ at 
the primary vertex. One of them decays into a $\mu^+$ after a path length of 1330$\mu$m. The kink angle is 209mrad and 
the Pt of the daughter particle is about 460MeV/c. The beginning of a photon electromagnetic shower is also seen.

\begin{figure}[h]
\centering
\includegraphics[width=70mm]{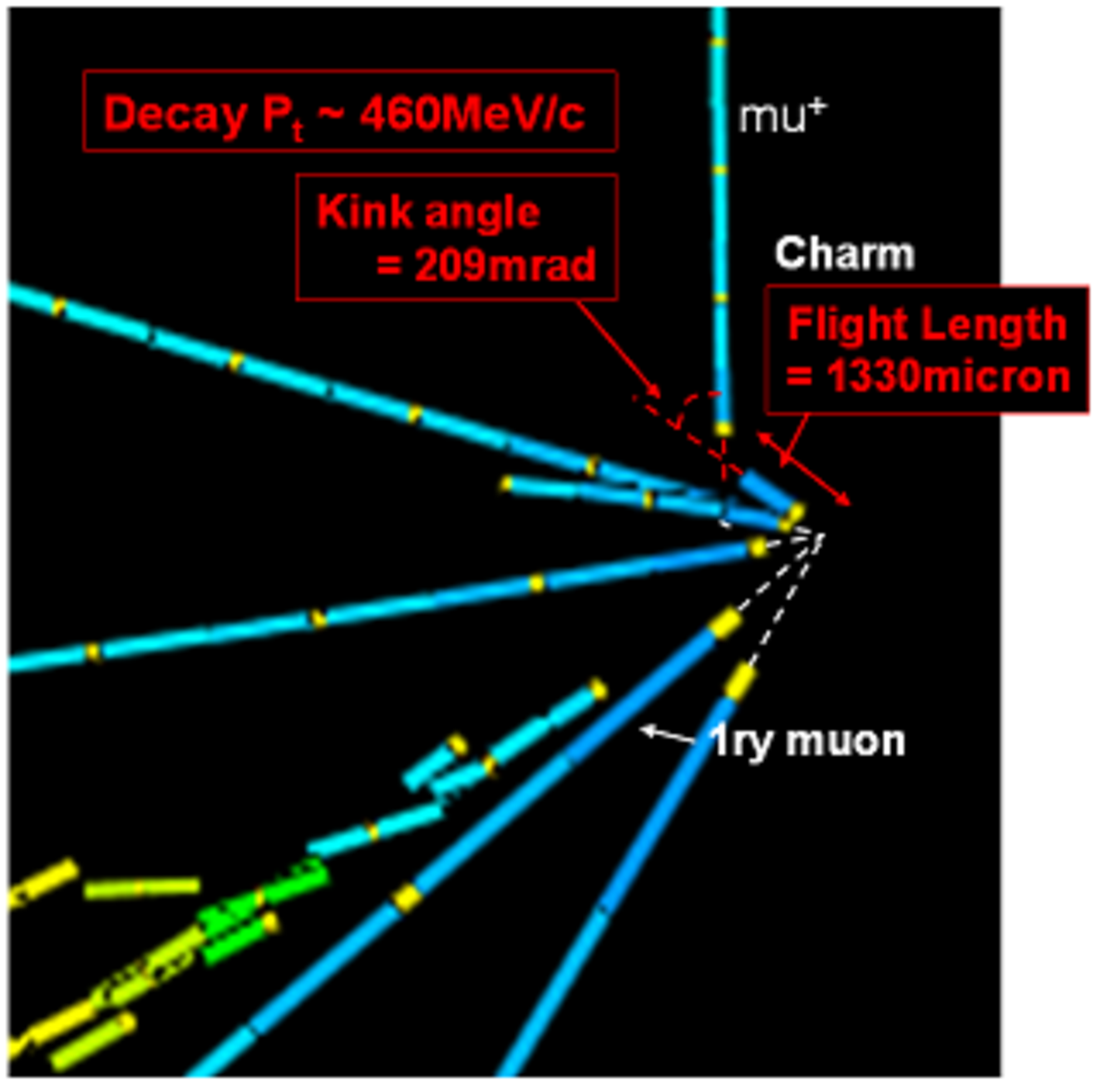}
\caption{A charm candidate event.} \label{figure16}
\end{figure}

\section{The 2009 OPERA Run}
The OPERA detector suffered no damage following the very strong L'Aquila earthquake of April 2009. The beginning of the 
run started on June 1 with a delay of a couple of weeks only. A nominal number of 4.5$\times$10$^1$$^9$ pot is requested 
from CERN, in which case about 4800 neutrino events would be collected in the OPERA targets by the end of the run and 
about 2.5 $\nu_\tau$CC interactions would be expected to be detected.

\section{Conclusions}
The goal of OPERA is to detect $\nu_\mu$ $\to$ $\nu_\tau$ oscillation in the appearance mode in the CERN CNGS beam. 
An innovative nuclear emulsion technology was developed for OPERA. In 2008, triggers for 1690 neutrino events in the 
targets were recorded. So far, the primary vertices of 797 of these events were located and analysed in the ECC bricks 
by a new generation of automatic scanning systems and the work is in progress. First examples of charm decay candidates 
were found and their kinematical analysis performed. The 2009 run has started on June 1 2009. It should lead to the 
potential detection of a couple of $\nu_\tau$CC interactions.


\bigskip 

\end{document}